\documentclass[twocolumn,preprintnumbers,amsmath,amssymb,superscriptaddress]{revtex4}
\usepackage{epsfig,bm}
\usepackage{ulem}
\usepackage{amsmath,color}

\begin{document}
\title{Neutron-skin thickness determines the
  surface tension of a compressible nuclear droplet}
\author{W. Horiuchi}
\affiliation{Department of Physics,
  Hokkaido University, Sapporo 060-0810, Japan}
\author{S. Ebata}
\affiliation{Nuclear Reaction Data Centre, Faculty of Science, 
Hokkaido University, Sapporo 060-0810, Japan}
\author{K. Iida}
\affiliation{Department of Mathematics and Physics, 
  Kochi University, Kochi 780-8520, Japan}

\begin{abstract}
  We systematically investigate the neutron-skin thickness of neutron-rich
  nuclei within a compressible droplet model, which includes
  several parameters characterizing the surface tension and the 
  equation of state (EOS) of asymmetric nuclear matter as well as
  corrections due to the surface diffuseness.  Such a systematic analysis
  helps towards constraining
  the EOS parameters of asymmetric nuclear matter and
  the poorly known density dependence of the surface tension; the latter
  is estimated
  with help of available experimental data
  for the neutron and proton density distributions and the nuclear masses.
  Validity of the present approach is confirmed by calculating
  realistic density distributions of Ca, Ni, Zr, Sn, Yb, and Pb isotopes 
  within a microscopic Skyrme-Hartree-Fock+BCS method for various
  sets of the effective nuclear force.  Our macroscopic model 
  accompanied by the diffuseness corrections works well in the sense
  that it well reproduces the evolution of the microscopically deduced
  neutron-skin thickness with respect to the neutron number for 
  selected sets of the effective nuclear force.  We find that
  the surface tension of the compressible nuclear droplet is a key to 
  bridging a gap between microscopic and macroscopic approaches.  
\end{abstract}
\maketitle

\section{Introduction}

Constraining the parameters that characterize the equation of state 
(EOS) of asymmetric nuclear matter from empirical data for laboratory 
nuclei is one of the possible approaches to the EOS.  However, 
the EOS parameters are still uncertain partly because
a significant fraction of nucleons lie in the surface region of a 
nucleus and partly because empirical data for neutron-rich unstable nuclei
are hard to obtain accurately.
For example, traditional electron elastic scattering measurements
have revealed that saturation of the density holds for 
stable nuclei.  Since experimental data are limited for 
such short-lived unstable nuclei, even the systematics of 
neutron and proton radii has not been established yet.
In fact, hundreds of theoretical models,
which equally well describe the saturation of the density and 
binding energy of stable nuclei, provide different EOS parameter sets 
(see, for example, Ref.~\cite{Dutra12}).

Astrophysical constraints on the EOS parameters might be relevant, e.g.,  
thanks to the recent precise mass measurement of a massive neutron 
star~\cite{Demorest10}, but such constraints would be influenced by
the poorly known EOS of neutron star matter at supranuclear densities.
It is thus still reasonable to focus on nuclear observables that are
sensitive to the symmetry energy at subnuclear densities.
One of such observables is neutron-skin thickness, which is defined by 
difference between root-mean-square (rms) point-neutron and 
point-proton radii.
In fact, a relationship between the symmetry energy and the 
neutron-skin thickness has actively been discussed
by several theoretical works~\cite{Chen10,Reinhard10,
  RocaMaza11,Kortelainen13,Inakura13,RocaMaza15}.

For stable nuclei, the charge-density distributions are well
determined from electron elastic scattering~\cite{deVries87}.
To determine the point-neutron radius experimentally, parity-violating 
electron scattering experiment was recently performed for 
$^{208}$Pb~\cite{PREX}. Uncertainty in the resultant neutron-skin thickness 
is relatively large, although further experiment is being planned 
towards higher precision~\cite{PREXII}. We thus focus on
proton elastic scattering experiments~\cite{Terashima08, Zenihiro10},
which allow one to extract the neutron radius 
from the overall fit of the differential cross sections 
up to the backward angles where the data are fairly sensitive 
to the elusive inner regime.
The experiments provide reliable
data for the neutron-skin thickness
of stable Pb and Sn isotopes.
For unstable nuclei, we remark that
the total reaction cross section
on a proton target has been used
to extract a neutron tail of halo nuclei
(see, for example, Refs.~\cite{Tanaka10, Moriguchi13}),
and is suggested as a promising tool to extract the neutron-skin 
thickness ~\cite{Horiuchi14,Horiuchi16}.
A combination of the total reaction and charge-changing
cross section measurements is also utilized
for this purpose with use of a carbon 
target~\cite{Estrade14,Terashima14,Suzuki16,Kanungo16}.

Theoretically many nonrelativistic and relativistic models 
for the effective nuclear force have been proposed in such a
way as to reproduce the saturation properties of stable nuclei, 
while each model corresponds to a particular set of the EOS 
parameters.  Classification of the models in terms of the EOS 
parameters is useful because such parameters are available for 
any form of the nuclear Hamiltonian.  Among others, the Skyrme
type Hamiltonian has more than hundreds of versions that give 
different sets of the EOS parameters through Skyrme-energy-density
functional (Skyrme-EDF) calculations~\cite{Dutra12}.
In Ref.~\cite{Brown13},  Brown selected sound Skyrme-EDF models
by making use of the neutron-skin thickness of doubly magic nuclei
as a constraint on the EOS parameters.  However, uncertainty in
the EOS parameters, particularly the slope parameter of the symmetry 
energy, $L$, is still large. 
As mentioned above, this comes from the fact that
many nucleons are present at around the nuclear surface.
It would thus be significant to consider
a relationship between the nuclear
  surface and the EOS parameters.

For this purpose, we take a macroscopic approach to the
neutron-skin thickness based on a compressible droplet 
model~\cite{Iida04}.  This model does not depend on details 
of the nuclear Hamiltonian but its underlying physics is the 
thermodynamics alone.  Traditionally, a nuclear droplet model is 
formulated by assuming that the droplet is incompressible,
but the nuclear density is not strictly saturated in 
finite nuclei. In fact, the nuclear droplet has to be 
compressible.  In the compressible droplet model, the surface 
tension depends generally on the density in the nuclear interior, 
while mechanical equilibrium determines the optimal value of
the internal density~\cite{Yamada64}.
In the case of neutron-rich nuclei in which 
nonzero neutron excess generally occurs in the nuclear interior 
even in the presence of the neutron skin, the optimal density 
in the interior is primarily controlled by $L$ through the 
saturation density of bulk matter that has the same neutron excess.
Then, the thermodynamics of the surface dictates the neutron-skin 
thickness of neutron-rich nuclei to have an explicit dependence 
on $L$ via the density dependence of the surface tension.
This is because 
the neutron skin, a manifestation of adsorption of excess neutrons 
onto the nuclear surface, is thermodynamically related to the shift 
of the surface tension due to a quasistatic change in the neutron 
excess in the nuclear interior.

Whereas the compressible droplet model roughly explains 
the neutron-skin thickness of stable nuclei, corrections 
that originate from surface diffuseness of the nuclei 
should be carefully taken into account to extract the bulk 
properties of nuclear matter.  Generally, such corrections, 
i.e., the surface width difference between neutrons and 
protons, are not considered although they can have 
nonnegligible effect~\cite{Warda09}. When one considers 
more neutron-rich nuclei, the effect has to be 
more significant because difference in the Fermi level
between protons and neutrons becomes larger.  Since the 
density dependence of the surface tension of the nuclear 
droplet is poorly known, furthermore, theoretical 
uncertainties are too large to constrain the EOS 
parameters~\cite{Iida04}.

In this paper, we revisit expression for neutron-skin 
thickness within a compressible droplet model proposed 
in Ref.~\cite{Iida04} and extend it by adding
surface diffuseness corrections between neutrons and protons.
For several sets of the EOS parameters that correspond to
the Skyrme effective interactions adopted here,
we determine the density 
dependence of the surface tension of the nuclear droplet 
in such a way that the expression for the neutron-skin
thickness is consistent with empirical data for the neutron 
and proton density distributions of stable Sn and Pb isotopes.
We then utilize the microscopic Skyrme-EDF method to 
calculate realistic density distributions 
of Ca, Ni, Zr, Sn, Yb, and Pb isotopes, as well as the EOS
parameters.  We finally compare the neutron-skin
thickness of Ca--Pb isotopes that can be calculated from the
macroscopic expression by using the determined density 
dependence of the surface tension with the results directly
evaluated from the microscopic Skyrme-EDF calculations.  We find 
that whether they agree well with each other or not depends on 
the adopted effective nuclear force.  This result opens a 
way to further constrain the EOS parameters.

In the next section, we give definitions of various quantities
of interest and brief explanations of our macroscopic 
models.  Section~\ref{cdroplet.sec} briefly explains a 
compressible droplet model.  A relationship between the neutron-skin 
thickness and the EOS parameters is also given
in terms of a primary factor that 
characterizes the neutron excess dependence of the neutron skin 
thickness.
In Sec.~\ref{surfcorr.sec}, nuclear surface width correction
to the droplet model expression for the neutron-skin thickness
is introduced.  We carefully define the nuclear surface width
or diffuseness for general nuclear density distributions
and use it for the correction.
Section~\ref{results.sec} presents our results and discussions.
After brief explanation of how we obtain
  realistic density distributions by a microscopic nuclear mean-field model
  in Sec.~\ref{density.sec}, we present, in Sec.~\ref{diffuseness.sec}, 
the surface widths, which are obtained from 
realistic nuclear densities, as what effectively
describe the surface properties
of neutrons and protons.
Then, in Sec.~\ref{factor.sec}, 
we determine the primary factor 
of the neutron-skin thickness
in the droplet model by using available experimental data.
This factor is correlated with
the parameter $\chi$ that controls the density dependence 
of the surface tension of the nuclear droplet.
Finally, a comparison of the microscopic theory and macroscopic
droplet model is made in terms of evolution of the neutron-skin 
thickness with respect to neutron excess in Sec.~\ref{ns.sec}.
Some microscopic models are 
not thermodynamically favored
because they fail in 
reproducing such evolution obtained by the macroscopic model and thus
do not satisfy the thermodynamic properties of finite 
nuclear matter.  Effects of the pairing interaction on
the nuclear surface are also discussed in Sec.~\ref{pairing.sec}.
Conclusions are given in Sec.~\ref{conclusions.sec}.

\section{Models}
\label{models.sec}

In this section we summarize basic features of our compressible
droplet model for nuclei and apply it to description of the neutron-skin 
thickness.  We then add corrections due to the surface diffuseness.

\subsection{Neutron-skin thickness in a compressible droplet model}
\label{cdroplet.sec}

\subsubsection{Definitions}

Let us consider an atomic nucleus, i.e., an $A$-nucleon system 
that consists of $N$ neutrons and $Z$ protons.  Neutron-skin thickness 
of this system is defined as difference between
point-neutron and proton rms radii:
\begin{align}
\Delta r_{np}=\left<r_n^2\right>^{\frac{1}{2}}-\left<r_p^2\right>^{\frac{1}{2}}.
\end{align}
These rms radii can be calculated by using the 
corresponding density distributions, $\rho_q(\bm{r})$, as
\begin{align}
  \left<r_q^2\right>=\frac{\int d\bm{r}\, r^2\rho_q(\bm{r})}
 {\int d\bm{r}\, \rho_q(\bm{r})},
\label{dens.eq}
\end{align}
where the subscript $q$ takes $p$ and $n$ for protons and 
neutrons, respectively.  It is noted that in the case of a sphere with 
uniform density distribution, the sharp cutoff radius, $R_q$, is 
related to the rms radius by
\begin{align}
R_q=\sqrt\frac{5}{3}\left<r_q^2\right>^{\frac{1}{2}}.
\label{rad.eq}
\end{align}
The point-nucleon (matter) rms radius is defined by 
\begin{align}
  \left<r^2\right>{^\frac{1}{2}}=
  \left(\frac{N}{A}\left<r^2_n\right>+\frac{Z}{A}\left<r_p^2\right>
  \right)^\frac{1}{2}.
\end{align}
We can use the same definitions as given in
Eqs.~(\ref{dens.eq}) and (\ref{rad.eq})
for the matter radius and density, but we omit the subscript $m$ for the 
sake of simplicity.  As a measure of neutron excess, it is convenient 
to define the asymmetry parameter:
\begin{align}
\delta=\frac{\rho_n-\rho_p}{\rho_n+\rho_p}.
\end{align}
Generally, $\delta$ is a function of $\bm{r}$, but we shall often 
take it as constant $\simeq (N-Z)/A$.
This approximation is good when $R_n\simeq R_p$.

\subsubsection{Compressible droplet model}

We now give expression for the neutron-skin thickness
in a compressible droplet model following Ref.~\cite{Iida04}.
In this model, a nucleus is viewed as a spherical liquid drop 
of variable uniform density $\rho_q$ and sharp cutoff radius $R_q$.
For nearly symmetric nuclei, which satisfies $R_n\simeq R_p$,
one can ignore the neutron-skin thickness at first approximation.
Then, the volume energy is $A$ times the bulk energy per nucleon, $w$, 
which can be expressed in a form expanded with respect to 
the matter density and neutron excess around $\rho=\rho_0$ and 
$\delta=0$~\cite{Lattimer81}:
\begin{align}
  w(\rho,\delta)&=w_0+\frac{K_0}{18\rho_0^2}(\rho-\rho_0)^2\notag\\
  &+\left[S_0+\frac{L}{3\rho_0}(\rho-\rho_0)\right]\delta^2,
\label{enervol.eq}
\end{align}
where $\rho_0$ and $w_0$ are the saturation density and the energy 
of symmetric nuclear matter. $K_0$, $S_0$, and $L$ are the 
so-called incompressibility of symmetric nuclear matter, 
the symmetry energy coefficient, and the density symmetry coefficient 
or slope parameter, respectively.  
Note that the saturation density of nearly
symmetric nuclear matter can be obtained from Eq.\ (\ref{enervol.eq})
as $\rho_0(1-3L\delta^2/K_0)$.  The surface energy is 
controlled by the density-dependent surface tension, $\sigma$, 
which can also be expanded as
\begin{align}
  \sigma(\rho,\delta)&=
  \sigma_0\left(1-C_{\rm sym}\delta^2+\frac{\chi}{\rho_0}(\rho-\rho_0)\right),
\label{enersurf.eq}
\end{align}
where $\sigma_0$ is the surface tension at $\rho=\rho_0$ and $\delta=0$, and
$C_{\rm sym}$ is the surface symmetry energy coefficient.
The parameter $\chi$ represents the density dependence
of the surface tension defined by
\begin{align}
    \chi\equiv\frac{\rho_0}{\sigma_0}
    \left.\frac{\partial \sigma}{\partial \rho}\right|_{\rho=\rho_0, \delta=0}.
\end{align}

In the compressible droplet model~\cite{Iida04}, a neutron
skin arises from adsorption of excess neutrons onto the surface, which 
is in turn in thermodynamic equilibrium with the bulk system of $A$ 
nucleons.  By separating the bulk system into the skin and interior
(neutron reservoir) regions, one can relate
the neutron-skin thickness with the EOS and surface parameters 
introduced in Eqs.\ (\ref{enervol.eq}) and (\ref{enersurf.eq}),
respectively.  For a given $R_p$, the neutron-skin thickness
can be expressed up to leading order in $\delta$ by
\begin{align}
  \Delta r_{np}^{\rm vol}\simeq \sqrt\frac{3}{5}
  \left[C\left(\delta-\frac{Ze^2}{20R_pS_0}\right)
    \left(1+\frac{3C}{2R_p}\right)^{-1}-\frac{Ze^2}{70S_0}\right]
\label{skinvol.eq}
\end{align}
with a primary factor
\begin{align}
C=\frac{2\sigma_0}{S_0\rho_0}\left(C_{\rm sym}+\frac{3L\chi}{K_0}\right).
\label{prifact.eq}
\end{align}
Note that the depression of the neutron-skin thickness due to the 
Coulomb interaction is considered in the formula by the terms 
involving $Z$.  The density dependence of the surface tension, 
$\chi$, which is a key parameter of this work, is correlated
with $L$ and $K_0$ as well as $C_{\rm sym}$.  So far the $\chi$ value is 
poorly known, but typically two values of $\chi$ are
assumed: $\chi=0$ in the absence of the density 
dependence~\cite{Myers69} and $\chi=4/3$ in the Fermi-gas model~\cite{Hilf66}.

\subsection{Diffuseness correction to neutron-skin thickness}
\label{surfcorr.sec}

Since the nuclear surface distribution is in general different 
for protons and neutrons, the surface width correction to the 
neutron-skin thickness occurs as the following term~\cite{Warda09}
\begin{align}
  \Delta r_{np}^{\rm surf}\simeq \sqrt\frac{3}{5}\frac{5}{2R}(b_n^2-b_p^2),
\label{skincorr.eq}
\end{align}
where $b_n$ ($b_p$) is the surface width of the neutron (proton) density.
If the density profile is the Fermi-type distribution,
$f(r)=(1+\exp[(r-\bar{R}_q)/a_q])^{-1}$,
the quantity $b_q$ can be related to the diffuseness parameter
$a_q$ by $b_q\sim \pi a_q/\sqrt{3}$.
The $b_n$ and $b_p$ values are typically taken as $\sim$ 1\,fm, 
which corresponds to the empirical diffuseness value of
$\sim 0.54$\,fm~\cite{BM}.

In this work, we employ realistic density distributions that can be
generated by a microscopic mean-field model, while we need a sound 
way of quantifying the surface width.  Warda {\it et al.}\ introduced 
a convenient definition of the surface width for one-dimensional 
half-infinite nuclear matter in equilibrium with the 
vacuum~\cite{Warda09}.  Here we extend it to a three-dimensional finite 
nucleus.  With a spherical density distribution, $\rho_q(r)$, and its 
derivative, $\rho^\prime_q(r)$, we calculate the mean location of 
the surface, $c_q$, by
\begin{align}
  c_q=\frac{4\pi\int_0^\infty r^3\rho_q^\prime (r)dr}{4\pi\int_0^\infty r^2\rho_q^\prime (r)dr}.
\end{align}
The square of the surface width can then be evaluated
from the mean-square radius of the gradient of the density 
distribution measured with reference to $c_q$ as
\begin{align}
  b_q^2&=\frac{4\pi\int_0^\infty (r-c_q)^2r^2\rho^\prime_q (r)dr}
{4\pi\int_0^\infty r^2\rho^\prime_q (r)dr}.
\label{surfwidth1.eq}
\end{align}
This definition is reasonable if the Fermi distribution well approximates 
$\rho_q(r)$.  This is because $b_q$ defined in Eq.\ 
(\ref{surfwidth1.eq}) approaches $\pi a_q/\sqrt{3}$ for large radius 
parameter $\bar{R}_q$ when the Fermi distribution is employed.
We remark that in the case of the trapezoidal distribution with the 
top-bottom length difference of $D_q$, it approaches $D_q/2\sqrt3$,
which is significantly small for the empirical value
of $D_q$ of order 2.2 fm~\cite{Kohama16}.

The Fermi distribution always gives an almost uniform
distribution in the interior region of a nucleus,
whereas any realistic density distribution exhibits some 
oscillatory behavior.  The derivative of such density distribution 
also oscillates and is not always small in the interior region.
In some cases, therefore, $b_q$ does not properly reflect the 
surface width or surface diffuseness, but it contains appreciable 
effects coming, e.g., from the internal depression of the density.
To avoid this problem, we assume that the surface diffuseness
is symmetric at $r=c_q$ and employ only the outer region of 
the integrand:
\begin{align}
  b_q^2&=\frac{2\int_{c_q}^\infty (r-c_q)^2[(r-c_q)^2+c_q^2]\,\rho^\prime_q (r)dr}
  {\int_{0}^\infty r^2\rho^\prime_q (r)dr}.
 \label{surfwidth.eq}
\end{align}
Note that the second term in the expansion of $r^2=(r-c_q)^2+2c_q(r-c_q)+c_q^2$
is omitted by assuming that $\rho^\prime_q$ is symmetric with respect to
$r=c_q$.
When we adopt Eq.\ (\ref{surfwidth.eq}) for realistic 
density distributions, as expected, the values of $b_q$ lie mostly 
between those of the trapezoidal and Fermi-type distributions.
Hereafter we shall thus use the above definition 
for $b_q$ unless otherwise mentioned.  Note that Eq.\ (\ref{skincorr.eq}) 
is derived by assuming that the nuclear matter has a flat interface
as was formulated in Ref.~\cite{Warda09}
but this $b_q$ defined here is for a three-dimensional density
distribution that includes the curvature effect of the nuclear sphere.
Since the effect is of higher order in the droplet model,
we can ignore this difference for medium-heavy and heavy nuclei.

\section{Results and discussions}
\label{results.sec}

In this section we present our calculations of the nuclear
surface diffuseness based on the microscopic theory,
determine the primary factor (\ref{prifact.eq})
that is consistent with empirical data for the neutron
and proton distributions, check the consistency 
between the microscopic and macroscopic evaluations of the neutron-skin
thickness, and finally discuss the pairing effect on the
nuclear surface.

\subsection{Density distributions
    with microscopic mean-field theory}
\label{density.sec}

Realistic density distributions of Ca, Ni, Zr, Sn, Yb, and Pb
isotopes are generated by the Skyrme-Hartree-Fock (HF) + BCS 
method in the three-dimensional coordinate space.
We employ a constant monopole pairing
as detailed in Refs.~\cite{Ebata14,Ebata17}.
All details of the calculation are given in Refs.~\cite{Ebata10, Horiuchi16}.
Since we do not assume any spatial symmetry in the calculation,
the deformation effect, which changes the structure of the
nuclear surface, is fully taken into account.
The obtained intrinsic density is generally deformed,
while the density distribution in the ground state 
is spherical in the laboratory frame.  Such a spherical density 
distribution can be obtained by taking the 
angle average as was done in Ref.~\cite{Horiuchi12}.  Validity 
of the resulting density distributions can be confirmed
by comparison with experimental data in the following 
way.  These density distributions, once build into an appropriate 
reaction theory based on the Glauber formalism~\cite{Glauber}, 
reproduce the total reaction cross sections~\cite{Horiuchi12,Horiuchi15}
obtained by the recent measurements~\cite{Takechi10,Takechi14}
within error bars.

The nuclear structure is somewhat sensitive to the Skyrme 
interaction employed.  For example, the SkM* and SLy4 interactions 
give a different neutron number dependence of the nuclear 
radii since nuclear deformations change 
the density profiles at around the nuclear surface
in a different manner~\cite{Horiuchi12}.
To test the interaction model dependence, we employ 
SkM*~\cite{SkMs}, SLy4~\cite{SLy4}, SkI3~\cite{SkI3}, 
KDE0v1~\cite{KDE0v1}, LNS~\cite{LNS}, SkT1,2,3~\cite{SkT123},
and SV-sym32~\cite{SV-sym32}. These Skyrme interactions except for
SkI3 belong to those selected
according to the classification suggested in Ref.~\cite{Brown13}.

\begin{figure}[ht]
  \begin{center}
\epsfig{file=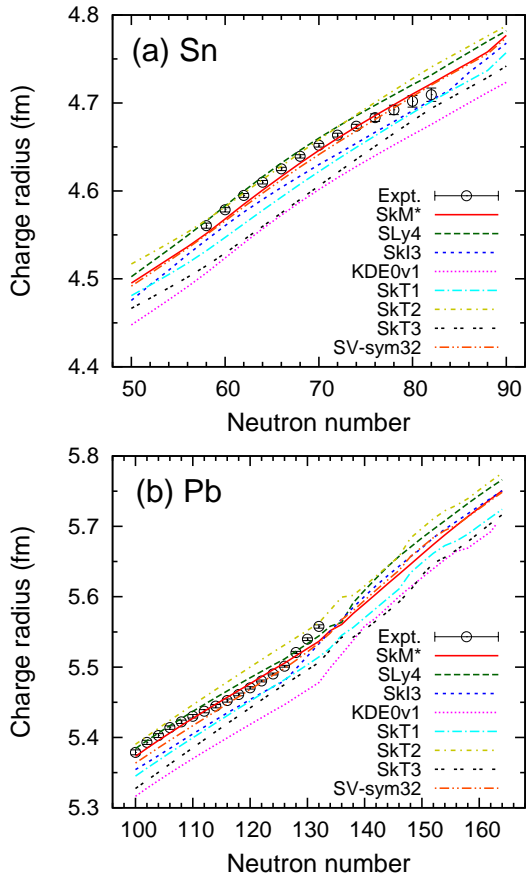, scale=1.2}        
\caption{(Color online) Charge radii of (a) Sn and (b) Pb isotopes.
  The point-proton radii obtained by the HF+BCS proton density distributions
  are converted to the charge radius by taking into account
  finite size corrections of the proton
  and neutron charge radii and the Darwin-Foldy term~\cite{Friar97}.
 The experimental charge radii are taken from Ref.~\cite{Angeli13}.}

\label{charge.fig}
  \end{center}      
\end{figure}

  To see the characteristics of the interactions,
  we compare, in Fig.~\ref{charge.fig},
  the calculated charge radii of Sn and Pb isotopes with
  experimental data~\cite{Angeli13}.
  The results with the LNS interaction are not plotted
  because the calculated charge radii are considerably smaller than
  those obtained with the other interactions by $\sim 0.1$\,fm.
  Though there are some quantitative differences,
  the results with the SkM*, SLy4, SkT2, and SV-sym32 interactions
  exhibit a fairly good agreement with the experimental charge radii,
  while the results with the SkI3, KDE0v1, SkT1, and SkT3 interactions
  deviate appreciably from the measured values.
  We remark that the SkI series which simulates the spin-orbit 
  strength of the relativistic mean-field model can reproduce the 
  kink~\cite{Goddard13}.  In fact, the SkI3 interaction alone shows 
  such a kink behavior at the neutron number 126 of Pb isotopes.
  
\subsection{Nuclear surface width}
\label{diffuseness.sec}

The nuclear surface has important information on the nuclear structure,
such as deformation, skin, and weakly-bound neutron orbits, etc.
In fact, the nuclear deformation changes density profiles
at the nuclear surface and enhances the nuclear size.  
Comparison of theoretical models
with the total reaction cross section measurements supports 
strong deformations in the neutron-rich Ne and Mg isotopes
~\cite{Minomo11,Minomo12,Sumi12,Horiuchi12,Watanabe14},
although whether or not this conclusion holds for 
any collision energy has yet to be clarified. 
The low-lying electric dipole ($E1$) strength
is also sensitive to the nuclear surface.
The abrupt change of the low-lying $E1$ strength 
at the magic numbers, which is possibly measured
by the total reaction cross section with a heavy 
target nucleus~\cite{Horiuchi17}, can be explained by
the structure change of the outermost single-particle 
orbit~\cite{Ebata14}.

\begin{figure}[ht]
  \begin{center}
\epsfig{file=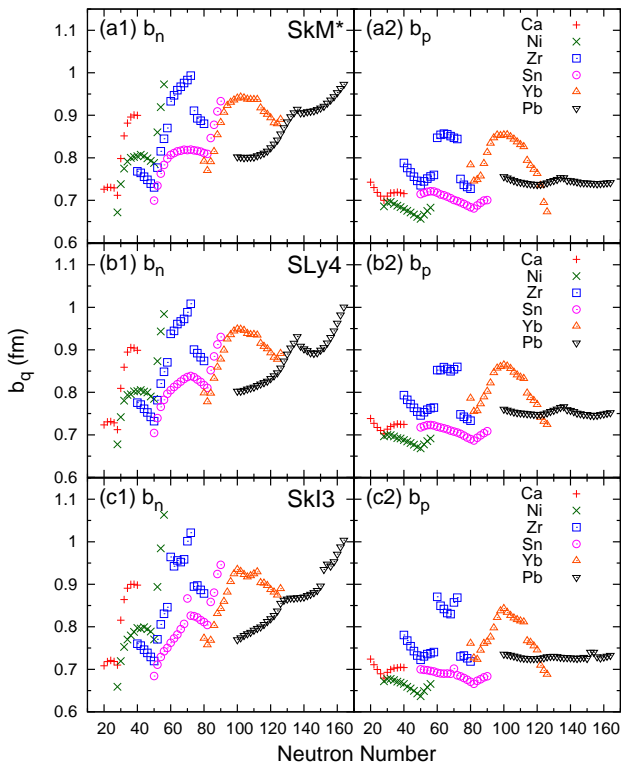, scale=0.9}        
\caption{(Color online) Surface widths
        of Ca, Ni, Zr, Sn, Yb, and Pb isotopes extracted from the
        (left) neutron and (right) proton density distributions,
        respectively.  For calculations of these distributions,
        the (a) SkM*, (b) SLy4,
          and (c) SkI3 interactions are employed 
        from top to bottom, respectively.}
  \label{diffHFBCS.fig}
  \end{center}      
  \end{figure}

It is interesting to see a systematic trend of the nuclear 
surface width of the HF+BCS densities.
Figure~\ref{diffHFBCS.fig} displays how the
surface widths obtained for the proton and neutron
densities of Ca, Ni, Zr, Sn, Yb, and Pb isotopes
depend on the neutron number.  These surface widths
are found to range between $\sim$0.6 and $\sim$1\,fm.  
Although there are some quantitative differences, 
all the Skyrme interactions show a similar neutron 
number dependence of the surface width.
We remark that the value of $b_q$ is generally smaller in the
Thomas-Fermi calculations~\cite{Warda09} in which the surface
diffuseness tends to be underestimated~\cite{Oyamatsu03}.

Since the surface width is closely related to
  the diffuseness of the nuclear surface,
the behavior of $b_q$
exhibits some interesting nuclear structure 
properties. Generally, 
$b_n$ increases as the Fermi level rises 
which allows the outermost neutron orbit to extend 
and hence gives larger diffuseness at the nuclear surface.
The behavior of $b_q$ is different in a way that
 depends on the quantum number, particularly, the angular 
 momentum of the outermost neutron orbits.
  In fact, sudden rises
are found at the spherical magic numbers, i.e., 
$N=28$ for Ca, $N=50$ for Ni, $N=82$ for Sn, and $N=126$ 
for Pb.  A change of the major shell or angular momentum of
the outermost single-particle orbit can be seen in $b_n$ at
the magic numbers. For Zr and Yb isotopes, the $b_n$ becomes
maximum in the open-shell regime between the magic numbers
because the density distribution at the nuclear surface extends
due to the nuclear deformation.

It is interesting to note that $b_p$ tends to decrease with 
neutron number, a tendency that stems from the fact that in 
general, the proton Fermi level becomes deeper with 
increasing neutron number.  Thus, in neutron-rich unstable 
nuclei, the proton density distribution at the nuclear surface 
is significantly sharp as compared with the neutron one. 
The local maxima and minima of $b_p$ arise basically 
by following the behavior of $b_n$.  This is natural 
because the interaction between protons and neutrons 
is strongly attractive and hence they tend to be 
close to each other.

\subsection{Primary factor that expresses the neutron-skin thickness
  in a compressible droplet model}
\label{factor.sec}

Here we propose a way to determine the unknown parameter
in the compressible droplet model from the existing experimental data.
In fact, the precise measurements of intermediate energy,
proton-elastic scattering cross sections have been performed for 
$^{116,118,120,122,124}$Sn~\cite{Terashima08} and 
$^{204,206,208}$Pb~\cite{Zenihiro10}.  By combining these measurements
with the proton density distributions extracted
from the electron-scattering measurements~\cite{deVries87},
the neutron density distributions and thus the neutron-skin
thickness have been extracted.

Within our model, the neutron skin thickness can be 
expressed as a sum of the volume and surface terms
given by Eqs.~(\ref{skinvol.eq}) and (\ref{skincorr.eq}):
$\Delta r_{np}=\Delta r_{np}^{\rm vol}+\Delta r_{np}^{\rm surf}$.
We determine $R_p$, $R$, $b_n$, and $b_p$ from the empirical
proton and neutron density distributions~\cite{Terashima08,Zenihiro10}
and then substitute the resultant values into the expressions for 
$\Delta r_{np}^{\rm vol}$ and $\Delta r_{np}^{\rm surf}$.  Aside from
$S_0$, which will be discussed just below, the volume term still 
contains one unknown factor, namely, $C$, 
Eq.\ (\ref{prifact.eq}), which roughly determines a slope of 
$\Delta r_{np}^{\rm vol}$ with respect to $\delta$.
We can thus fix the factor $C$ in such a way as to 
minimize the rms deviation from the empirical neutron-skin 
thickness as defined by
$\sqrt{\frac{1}{N_d}\sum_{i=1}^{N_d}[\Delta r_{np}(i)-\Delta r_{np}^{\rm Expt.}(i)]^2}$,
where $N_d$ denotes the number of available data.

\begin{figure}[ht]
\begin{center}
  \epsfig{file=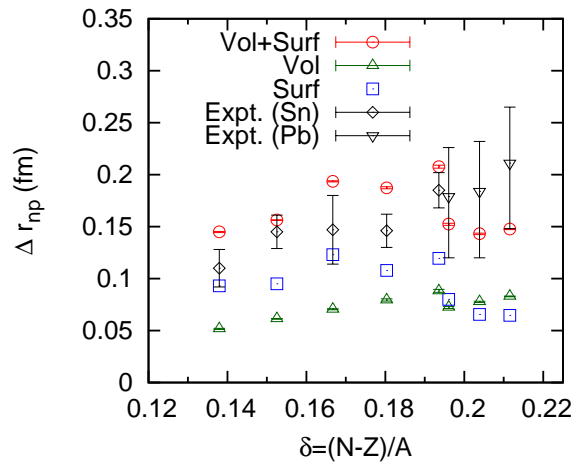, scale=1.4}        
  \caption{(Color online) Neutron-skin thickness 
    (circles) calculated for Sn and Pb
    isotopes within the compressible droplet model
    using the empirical 
    density distributions~\cite{Terashima08,Zenihiro10}.
    Decomposition into the volume (triangles) and surface 
    (squares) terms and the empirical skin-thickness 
    values (inverted triangles for Pb; diamonds for Sn) are also given.
    Error bars of the total neutron-skin thickness
      and its volume contribution indicate
      a range of the calculated values with $S_0=28$--36\,MeV.
}
  \label{skincomp.fig}
  \end{center}      
\end{figure}

Figure~\ref{skincomp.fig} displays the theoretical and experimental
$\Delta r_{np}$ for stable Sn and Pb isotopes.
Given phenomenological estimates of the symmetry energy coefficient,
$S_0=32\pm 4$\,MeV~\cite{Oyamatsu03}, we obtain 
  $C=1.06^{+0.08}_{-0.07}$
and find that uncertainty in $C$ that comes from the error
$\pm4$\,MeV of $S_0$ is much smaller than that from the 
experimental uncertainty~\cite{Terashima08,Zenihiro10}.
This means that the results for $C$ are dictated by the
measurements, irrespective of the assumed values of the EOS 
and surface parameters.
  We neglect uncertainties in the surface term,
   which come partly from unpublished uncertainties in
    the deduced neutron and proton density distributions.
Another factor is
the shell and pairing effects, which modify the nuclear 
surface profile and are effectively included in the surface term of 
Eq.~(\ref{skincorr.eq}), as will be discussed in 
Sec.~\ref{pairing.sec}.  Judging from Fig.\ \ref{skincomp.fig},
however, we note that those effects
have to be also included in the volume term or $C$ 
in such a way that $C$ is smaller (larger)
for Sn (Pb) isotopes than the above value,
but remain to be examined in the present qualitative analysis.
For better estimates of $C$,
it would be significant to increase the number of empirical
data for the proton and neutron density distributions of 
neutron-rich nuclei.
In the next subsection, we will discuss the validity of 
the obtained $C$ by using the microscopic HF+BCS model
calculations.

\subsection{Comparison of the neutron-skin thickness 
between the macroscopic and microscopic models}
\label{ns.sec}

\begin{figure*}[ht]
  \begin{center}
    \epsfig{file=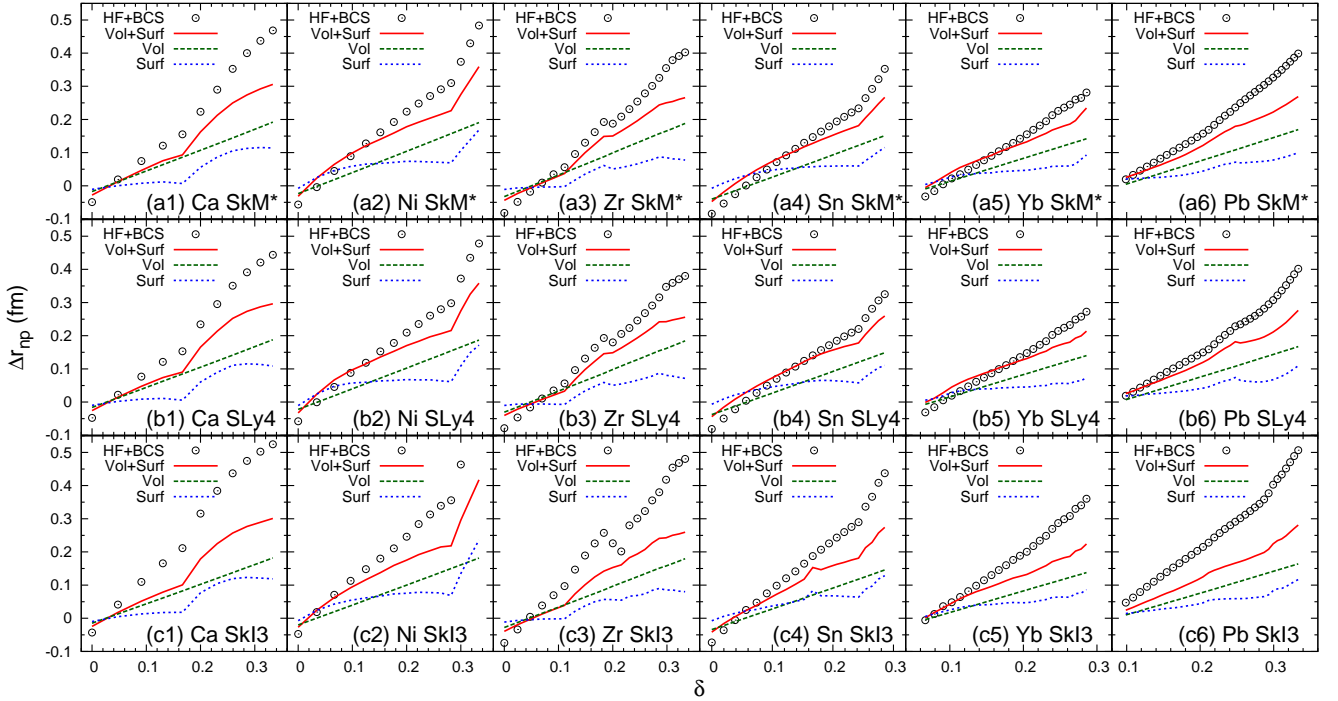, scale=0.83}
    \caption{(Color online)
      Neutron-skin thickness of Ca, Ni, Zr, Sn, Yb, and Pb isotopes
      (from left to right, respectively) as a function of the 
      asymmetry parameter $\delta=(N-Z)/A$.   
      The SkM*, SLy4, and SkI3 interactions (from top 
      to bottom, respectively) are employed for 
      the HF+BCS calculations (circles) and for the droplet formula 
      (solid lines) with the volume term (dashed lines) and 
      the surface term (dotted lines).
      }
\label{skinFermi.fig}
  \end{center}
\end{figure*}

Here we show usefulness of our macroscopic formula
for the neutron-skin thickness
by comparing it with the skin-thickness obtained 
by the Skyrme-HF+BCS model.  The EOS parameters predicted by 
hundreds of the Skyrme interactions within the HF framework
are available in Ref.~\cite{Dutra12}, where
for a given Skyrme parameter set, the corresponding EOS 
parameters are listed.

First, we redetermine $C$ by the same procedure as 
described in the previous subsection but for the $S_0$ 
value that corresponds to the given Skyrme-EDF.  
Although this value ranges approximately from 30 to 35\,MeV, 
the $S_0$ dependence of $C$ is tiny as shown above, and 
hence the redetermined value of $C$ lies in the range of $C$ 
as obtained above from stable Sn and Pb isotopes.
We can then fix the unknown parameter in the compressible droplet 
model (\ref{prifact.eq}), namely, the density dependence 
of the surface tension, $\chi$.
Here we set the values of $\sigma_0$ and $C_{\rm sym}$ 
to be 1\,MeV fm$^{-2}$ and 1.9, respectively, 
which are determined by the global fit of experimental 
nuclear masses within the framework of
the incompressible droplet model~\cite{Koura05}.  
This is reasonable because each effective interaction is 
constructed in such a way as to reproduce the same measured
masses of stable nuclei.  Note, however, that there are 
uncertainties in the above values of $\sigma_0$ and $C_{\rm sym}$.
Even in the incompressible limit, $C_{\rm sym}$ is uncertain
as will be shown below in the present subsection.  Once the effect of 
finite compressibility is included, furthermore, the global fit would 
redetermine $\sigma_0$ and $C_{\rm sym}$.  For simplicity, in the 
present qualitative analysis, we ignore such feedback corrections on 
$\sigma_0$ and $C_{\rm sym}$, which would have to be allowed for 
for more quantitative analysis.  We remark in passing that
$C_{\rm sym}=1.9$  is also consistent with the empirical 
$A$-dependence of the energy position of the giant dipole 
resonance~\cite{Trippa08}.

We can now utilize the optimal values of $C$ to examine
how well the compressible droplet model can reproduce the 
neutron-skin thickness calculated from the HF+BCS model 
for Ca, Ni, Zr, Sn, Yb, and Pb isotopes.  Here, 
we determine $R_p$, $R$, $b_n$, and $b_p$ from the HF+BCS
proton and neutron density distributions
and then substitute the resultant values, together with the
corresponding values of $C$ and $S_0$, into the expressions for
$\Delta r_{np}^{\rm vol}$ and $\Delta r_{np}^{\rm surf}$.
The results from the three Skyrme interactions, 
SkM*, SLy4, and SkI3, are shown in Fig.\ \ref{skinFermi.fig}.  
We also plot the decomposition of 
$\Delta r_{np}$ into the volume and surface terms in the droplet model.
The volume and surface contributions are found to be comparable for all 
the nuclides considered here.  The volume term monotonically 
increases almost linearly with $\delta$, whereas the surface term 
increases in such a way as to reflect the difference
of the surface widths or diffuseness of protons and neutrons.
As for SkM* and SLy4, not only large enhancement of 
$\Delta r_{np}$ due to weakly bound orbits beyond the neutron
magic numbers 28, 50, 82, and 126 for spherical Ca, Ni, Sn, 
and Pb isotopes, respectively, but also 
zigzag patterns for Zr isotopes, which stem from 
the nuclear deformation, are fairly well reproduced. 
Such reproduction of the local structure is ensured by 
the surface term.  It is to be noted that as far as the SkI3 
interaction is concerned, the droplet results for 
$\Delta r_{np}$ deviate considerably from the HF+BCS ones, 
for Pb isotopes in particular.

 For the SkM* and SLy4 interactions,
our macroscopic model fairly well reproduces $\Delta r_{np}$
obtained by the HF+BCS calculations up to $\delta \sim 0.2$.
Since the formula given by Eqs.~(\ref{skinvol.eq}) and 
(\ref{skincorr.eq}) assumes $R_p\sim R_n$,
higher order terms, which start with the quadratic 
term~\cite{Iida04}, should be considered for
more quantitative description of the regime
$\delta \gtrsim 0.2$.

\begin{figure}[ht]
  \begin{center}
    \epsfig{file=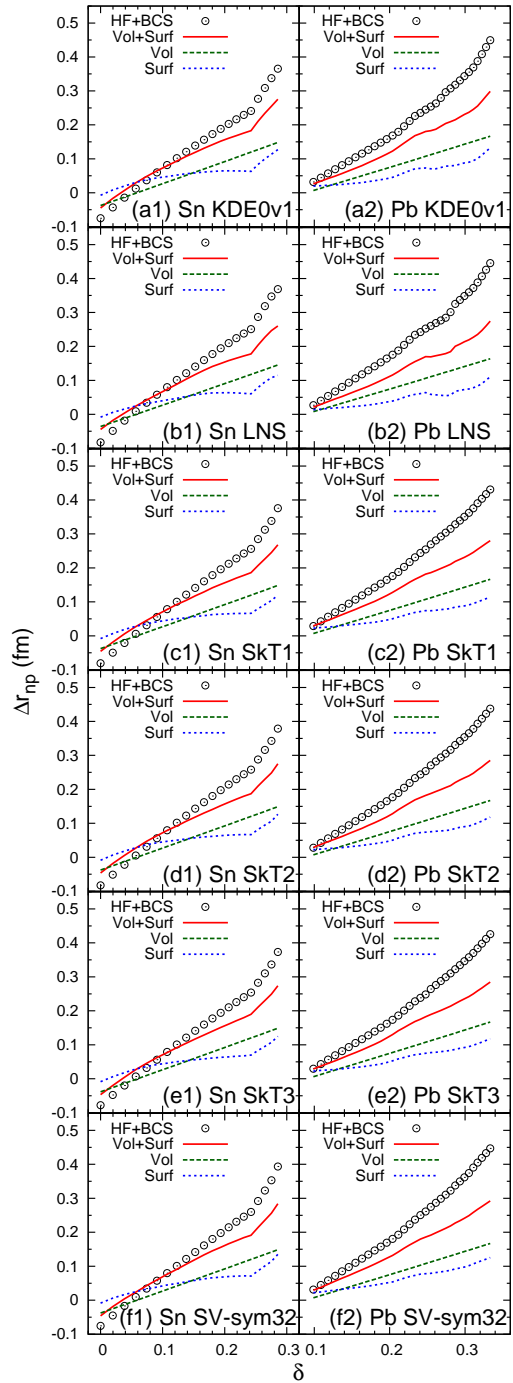, scale=0.85}
    \caption{(Color online) Same as Fig.~\ref{skinFermi.fig}
      but for (left) Sn and (right) Pb isotopes.
      with the (a) KDE0v1, (b) LNS, (c) SkT1, (d) SkT2, (e) SkT3,
        and (f) SV-sym32 interactions.}
\label{skinSnPb.fig}
  \end{center}
\end{figure}

To test the interaction dependence further,
we display, in Fig.~\ref{skinSnPb.fig}, the same plot as
Fig.~\ref{skinFermi.fig} but for Sn and Pb isotopes
with the KDE0v1, LNS, SkT1--3, and SV-sym32 interactions.
The KDE0v1 and SV-sym32 interactions show a 
marginally good agreement of the macroscopic result for
$\Delta r_{np}$ with the microscopic one
obtained by the HF+BCS calculations for Sn and Pb isotopes,
while the LNS exhibits an appreciable difference between these two.
The SkT1--3 interactions give a reasonable agreement, which
is better for Pb isotopes than that for Sn isotopes.

Table~\ref{eos.tab} summarizes the EOS parameters, which are 
taken from Ref.~\cite{Dutra12}, and the extracted $\chi$ values
for various sets of the Skyrme interactions.  Recall that
the $\chi$ is correlated with the surface symmetry coefficient 
$C_{\rm sym}$.   The smallest value of $C_{\rm sym}$ as assumed 
here is 1.4, which can be obtained from a 
Bethe-Weizs{\" a}ecker type mass formula that includes
the surface symmetry term.  Then, we set $C_{\rm sym}=1.9\pm 0.5$, 
which in turn determines uncertainly in $\chi$ 
given that $\sigma_0$ is known much better.
Note that the values of $\chi$ obtained for the SkM* and SLy4 
interactions are generally close to the Fermi-gas-model 
prediction $\chi=4/3$. As an exception, the SkI3 interaction 
gives a considerably smaller $\chi$, which reflects the fact that 
the corresponding EOS parameter $L/K_0$ is significantly larger 
than those of the other interactions.

Since $K_0$ is not strongly dependent on the Skyrme interaction employed,
it is interesting to  focus on the value of $L\chi$ for each of the 
Skyrme EDF models.  The $L\chi$ values are 50--70\,MeV for all
the interactions except for the SkI3 and LNS interactions.
These exceptional interactions give $L\chi > 70$\,MeV,
resulting in the poor reproduction of the microscopically obtained
$\Delta r_{np}$.
We remark that the EOS parameters of the SkI3 interaction
is excluded in the constraint with the neutron-skin thickness of
doubly-closed nuclei~\cite{Brown13} and also 
in the unitary gas constraint~\cite{Kolomeitsev17}.
It should be noted that
  our analysis is based on the specific model, namely, 
  nonrelativistic mean-field with the Skyrme effective interaction.
  Further investigation with other models, e.g. relativistic mean-field model,
  would be desired to further confirm whether the finding 
  obtained here is universal or not.

\begin{table}[h]
  \caption{EOS parameters derived from 9 Skyrme-EDF models~\cite{Dutra12}
    and the density dependence of the surface tension, $\chi$, that is
    consistent with the empirical $C$ value.
    Units are given in fm$^{-3}$ for $\rho_0$ and MeV for $S_0$, $K_0$, and $L$.}
    \begin{tabular}{ccccccc}
      \hline\hline
Name&&$\rho_0$ &$S_0$&$K_0$ &$L$ &$\chi$ ($C_{\rm sym}=1.9\mp 0.5$)\\
      \hline
SkM*    &&  0.160&30.03&216.61&45.78 &1.16$\pm$0.79\\
SLy4    &&  0.160&32.00&229.91&45.94 &1.35$\pm$0.83\\
SkI3    &&  0.158&34.83&258.19&100.53&0.76$\pm$0.43\\
KDE0v1  &&  0.165&31.97&223.90&41.42 &1.60$\pm$0.90\\
LNS     &&  0.175&33.43&210.78&61.45 &1.28$\pm$0.57\\
SkT1    &&  0.161&32.02&236.16&56.18 &1.15$\pm$0.70\\
SkT2    &&  0.161&32.00&235.73&56.16 &1.15$\pm$0.70\\
SkT3    &&  0.161&31.50&235.74&55.31 &1.14$\pm$0.71\\
SV-sym32&&  0.159&32.00&233.81&57.07 &1.08$\pm$0.68\\
\hline\hline
 \end{tabular}
\label{eos.tab}
\end{table}

\subsection{Effect of pairing on nuclear surface}
\label{pairing.sec}

Generally, the nuclear diffuseness reflects the 
structure around the nuclear surface, while
the pairing correlation
is known to play an essential role in
a realistic description of the nuclear surface
because it defines the occupations of single-particle states
  near the Fermi surface.
Here we discuss the effect of the pairing correlation
on the nuclear surface width
  of the density distributions obtained by the 
microscopic calculations.

\begin{figure}[ht]
  \begin{center}
    \epsfig{file=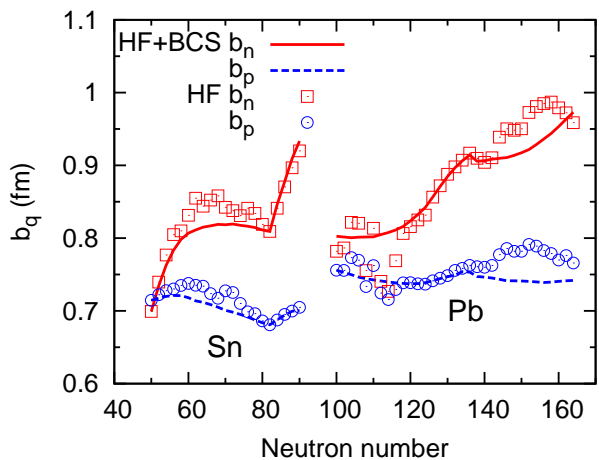, scale=1.4}
    \caption{(Color online)
      Surface widths of Sn and Pb
      isotopes that are extracted from the
      neutron and proton density distributions 
      calculated in the presence (HF+BCS) and 
      absence (HF) of the pairing interaction.
      The SkM* interaction is employed
      for calculations of the density distributions.}
    \label{diffHF.fig}
\end{center}    
\end{figure}

Figure~\ref{diffHF.fig} plots the surface widths
  obtained from the density distributions of Sn and Pb isotopes that are 
calculated in the presence and absence of the pairing interaction.  
The surface widths tend to be large
when the pairing interaction is ignored.
In most cases, the pairing correlation plays a role
in reducing the degree of the nuclear deformation.
In fact, in the presence of the pairing interaction, 
all the Sn isotopes have a spherical shape, whereas
the deformed ground states occur when
the pairing interaction is off, resulting in such an 
  artificial increase of the rms radii~\cite{Horiuchi16} as is 
  not seen in the experimental charge radii~\cite{Angeli13}.
At $N\geq 82$, on the other hand, a spherical shape 
is robust in both cases.
All the Pb isotopes
  also show a spherical shape in the presence of the paring interaction.
  The effect appears to be small at $120 \lesssim N \lesssim 140$
  in which the HF ground states have an almost spherical shape,
  while we see some difference at the neutron- and proton-rich regions
  where a large deformation appears in the HF ground states.

\begin{figure}[ht]
  \centering
    \epsfig{file=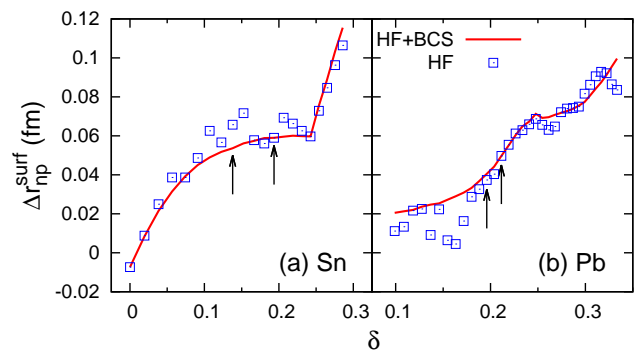, scale=1.05}
    \caption{(Color online)
      Same as Fig.\ \ref{diffHF.fig} but for the surface 
      contribution to the neutron-skin thickness of (a) Sn and (b) Pb isotopes.
      For the guide of eyes, arrows that indicate
        the region of $^{116-124}$Sn and $^{206-208}$Pb in which
        experimental data are available are drawn in
        the panels (a) and (b), respectively.
    }
\label{skincorr.fig}
\end{figure}

Figure~\ref{skincorr.fig} plots the surface correction term
of Eq.~(\ref{skincorr.eq}) calculated by allowing for and
ignoring the pairing interaction for Sn and Pb isotopes.
  Since both proton and neutron distributions are deformed,
  subtraction of the surface widths between protons and neutrons
 in Eq.\ (\ref{skincorr.eq})
somewhat cancels the effect of the nuclear deformation.
Although there is
no significant difference between both cases, switch-off
of the pairing interaction allows an artificial 
zigzag pattern, which is not seen in the experimental 
neutron-skin thickness~\cite{Terashima08, Zenihiro10},
to appear in the surface correction term at the open shell regions. 
This suggests that the pairing interaction plays an 
important role in correctly describing the nuclear surface.
 In order to extract the EOS parameters from finite nuclei, therefore,
 detailed study on the effect of the paring correlation will be 
 indispensable.  In this paper, we employ
 a constant-monopole-type paring~\cite{Ebata14,Ebata17}
 as one of the standard paring interactions.
 Investigations with other types of the paring interaction
 would be interesting but it is beyond the scope of this paper.

\section{Conclusions}
\label{conclusions.sec}

In summary, to revisit a relation between
the neutron skin thickness of finite nuclei and the 
EOS of asymmetric nuclear matter, we apply a compressible 
nuclear droplet model including an appropriate correction 
of the neutron and proton surfaces to description of 
the neutron-skin thickness.  This is a significant
update of the previous work~\cite{Iida04}, which has not
included any corrections due to the surface diffuseness.
For several sets of the EOS 
parameters that correspond to the specific
Skyrme effective interactions, 
we determine the density dependence of the surface tension $\chi$  
in the nuclear droplet from empirical data for the neutron 
and proton density distributions of stable Sn and Pb isotopes.
Such determination provides possible way
of determining  the density dependence of the surface tension, which is a key 
quantity to bridge a gap between microscopic and macroscopic 
nuclear models.

We also present a reasonable definition of the surface 
width of the nuclear density distribution by using
realistic density distributions of Ca, Ni, Zr, Sn, Yb, and Pb isotopes
that are generated by the microscopic
Skyrme Hartree-Fock (HF) + BCS model.
We confirm from our macroscopic model that
the difference of the proton and neutron 
surface widths plays a decisive role in determining 
the neutron-skin thickness.  In fact, the surface width 
correction to the thickness can be comparable to
the volume contribution, which contains
information on the bulk properties of nuclear matter.
This seems to be one of the reasons why the parameter $L$ 
characterizing the density dependence of the symmetry energy, 
which does have a strong correlation with 
the skin thickness, is still uncertain.

 Another reason for that could be uncertainties in the 
surface tension, which, together with the bulk properties,
controls the volume contribution.  Even with $\chi$ being 
determined in the present analysis, the surface symmetry 
coefficient $C_{\rm sym}$ has yet to be precisely
fixed by experimental data.

 Fortunately, we still have some chance of constraining the 
EOS parameters.  This is based on the consistency check of 
the thermodynamic droplet description of the neutron skin thickness 
with the HF+BCS prediction for each of the Skyrme interactions 
adopted here.
We find most of the Skyrme interactions
  have $\chi$ of the order of the Fermi-gas value 4/3.
  In particular, the SkM* and SLy4 interactions
  show an almost perfect consistency between
  the microscopically and macroscopically
 obtained neutron-skin thickness.
 A group with $L\chi\sim 50$-70\,MeV, in which
 the SkM* and SLy4 interactions are included,
 shows a good consistency in contrast to a group with $L\chi\gtrsim 70$\,MeV.
 This implies that the latter group is not thermodynamically 
favored, although a more quantitative 
analysis that allows for shell and pairing effects on the primary 
factor $C$ would be desired to make sure of that.

To obtain a better constraint on the EOS parameters,
systematic investigations on the surface diffuseness
of nuclei including neutron-rich unstable ones would be 
necessary. The surface width or diffuseness
of unstable nuclei 
could be experimentally determined, e.g., by using 
elastic scattering in inverse kinematics with a proton target.
Such measurements would be 
hopefully made in the near future to deepen our understanding 
of asymmetric nuclear matter. 

\acknowledgments

We thank S. Terashima for sending us
the numerical data of proton and neutron density distributions
of Sn isotopes. We are also grateful to J.~M. Lattimer
for valuable communications.
 Computational resources were in part provided
  by Research Center for Nuclear Physics, Osaka University.
This work was performed in part at
Aspen Center for Physics,  which is supported by the NSF grant
PHY-1607611.   Also, this work was supported in part by 
Grants-in-Aid for Scientific Research on Innovative Areas 
through No.\ 24105008 provided by MEXT.


\begin{thebibliography}{99}
\bibitem{Dutra12} M. Dutra, O. Louren\c{c}o, J.~S. S\'{a}\ Martins,
  A. Delfino, J.~R. Stone, and P.~D. Stevenson,
Phys.\ Rev.\ C {\bf 85}, 035201 (2012).
\bibitem{Demorest10}
  P.~B. Demorest, T. Pennucci, S.~M. Ransom, M.~S.~E. Roberts,
  and J.~W.~T. 
  Hessels, Nature (London) {\bf 467}, 1081 (2010).
\bibitem{Chen10}
L.~W. Chen, C.~M. Ko, B.~A. Li, and J. Xu,
Phys.\ Rev.\ C {\bf 82}, 024321 (2010).
\bibitem{Reinhard10}
P.-G. Reinhard and W. Nazarewicz,
Phys.\ Rev.\ C {\bf 81}, 051303(R) (2010).
\bibitem{RocaMaza11}
X. Roca-Maza, M. Centelles, X. Vi\~nas, and M. Warda,
Phys.\ Rev.\ Lett.\ {\bf 106}, 252501 (2011).
\bibitem{Kortelainen13} 
M. Kortelainen, J. Erler, W. Nazarewicz, N. Birge, Y. Gao, and E. Olsen,
Phys.\ Rev.\ C {\bf 88}, 031305(R) (2013).
\bibitem{Inakura13}
T. Inakura, T. Nakatsukasa, and K. Yabana,
Phys.\ Rev.\ C {\bf 88}, 051305(R) (2013).
\bibitem{RocaMaza15}
X. Roca-Maza, X. Vi\~nas, M. Centelles, B.~K. Agrawal,
G. Col\'o, N. Paar, J. Piekarewicz, and D. Vretenar,
Phys.\ Rev.\ C {\bf 92}, 064304 (2015).
\bibitem{deVries87} H. de Vries, C.~W. Jager, and C. de Vries,  
At.\ Data Nucl.\ Data Tables {\bf 36}, 495 (1987).
\bibitem{PREX} S. Abrahamyan {\it et al.},
  Phys.\ Rev.\ Lett.\ {\bf 108}, 112502 (2012).
\bibitem{PREXII} Proposal to Jefferson Lab PAC 38, available
  at http://hallaweb.jlab.org/parity/prex/prexII.pdf
\bibitem{Terashima08} S. Terashima {\it et al.},
Phys.\ Rev.\ C {\bf 77}, 024317 (2008).
\bibitem{Zenihiro10} J. Zenihiro {\it et al.},
  Phys.\ Rev.\ C {\bf 82}, 044611 (2010).
\bibitem{Tanaka10} K. Tanaka {\it et al.}, Phys. Rev. Lett.
  {\bf 104}, 062701 (2010). 
\bibitem{Moriguchi13} T. Moriguchi {\it et al.}, Phys. Rev. C {\bf 88},
    024610 (2013).    
\bibitem{Horiuchi14}
  W. Horiuchi, Y. Suzuki, and T. Inakura, Phys.\ Rev.\ C {\bf 89}, 011601(R) 
(2014).
\bibitem{Horiuchi16} W. Horiuchi, S. Hatakeyama, S. Ebata, and Y. Suzuki,
Phys.\ Rev.\ C {\bf 93}, 044611 (2016).
\bibitem{Estrade14} A. Estrad\'{e} {\it et al.},
  Phys.\ Rev.\ Lett.\ {\bf 113}, 132501 (2014).
\bibitem{Terashima14} S. Terashima {\it et al.},
Prog.\ Theor.\ Exp.\ Phys.\ {\bf 2014}, 101D02 (2014).
\bibitem{Suzuki16} Y. Suzuki {\it et al.}, Phys.\ Rev.\ C {\bf 94},
  011602(R) (2016).
\bibitem{Kanungo16} R. Kanungo {\it et al.}, Phys.\ Rev.\ Lett.\ {\bf 117}, 102501 (2016).
\bibitem{Brown13}
  B.~A. Brown, Phys.\ Rev.\ Lett.\ {\bf 111}, 232502 (2013).  
\bibitem{Iida04} K. Iida and K. Oyamatsu, Phys.\ Rev.\ C {\bf 69}, 037301 
(2004).
\bibitem{Yamada64} M. Yamada, Prog.\ Theor.\ Phys.\ {\bf 32}, 512 (1964).
\bibitem{Warda09} M. Warda, X. Vi\~nas, X. Roca-Maza, and M. Centelles,
  Phys.\ Rev.\ C {\bf 80}, 024316 (2009).
\bibitem{Lattimer81} J.~M. Lattimer, Ann.\ Rev.\ Nucl.\ Part.\ Sci.\ {\bf 31},
  337 (1981).  
\bibitem{Myers69} W.~D. Myers and W.~J. Swiatecki, Ann.\ Phys.\ (N.Y.) 
{\bf 55}, 395 (1969); {\it ibid.} {\bf 84}, 186 (1974). 
\bibitem{Hilf66} E. Hilf and G. S{\" u}ssmann, Phys.\ Lett.\ {\bf 21},
654 (1966). 
\bibitem{BM}
A. Bohr and B.~R. Mottelson, Nuclear Structure, Vol. I (W.~A. Benjamin,
New York, 1975).  
\bibitem{Kohama16} A. Kohama, K. Iida, and K. Oyamatsu,
  J. Phys.\ Soc.\ Jpn.\ {\bf 85}, 094201 (2016).
\bibitem{Ebata14}
S. Ebata, T. Nakatsukasa, and T. Inakura, Phys.\ Rev.\ C {\bf 90}, 024303 
(2014).
\bibitem{Ebata17} S. Ebata and T. Nakatsukasa,
  Phys.\ Scr.\ {\bf 92}, 064005 (2017).
\bibitem{Ebata10}
S. Ebata, T. Nakatsukasa, T. Inakura, K. Yoshida, Y. Hashimoto, 
and K. Yabana, Phys.\ Rev.\ C {\bf 82}, 034306 (2010).
\bibitem{Horiuchi12} W. Horiuchi, T. Inakura, T. Nakatsukasa, and Y. Suzuki,
  Phys.\ Rev.\ C {\bf 86}, 024614 (2012).
\bibitem{Glauber} R.~J. Glauber, {\it Lectures in Theoretical Physics}, 
edited by W.~E. Brittin and L.~G. Dunham (Interscience, New York, 1959), 
Vol.\ 1, p.315.  
\bibitem{Horiuchi15} W. Horiuchi, T. Inakura, T. Nakatsukasa, and Y. Suzuki,
 JPS Conf.\ Proc.\ {\bf 6}, 030079 (2015).
\bibitem{Takechi10} M. Takechi {\it et al.}, Mod.\ Phys.\ Lett.\ A {\bf 25}, 
1878 (2010).
\bibitem{Takechi14} M. Takechi {\it et al.}, Phys.\ Rev.\ C {\bf 90}, 
061305(R) (2014).  
\bibitem{SkMs}
J. Bartel, P. Quentin, M. Brack, C. Guet, and H. H{\aa}kansson, Nucl.\ Phys.\ 
{\bf A386}, 79 (1982).
\bibitem{SLy4} 
E. Chabanat, P. Bonche, P. Haensel, J. Meyer, and R. Schaeffer,
Nucl.\ Phys.\ A {\bf 627}, 710 (1997).
\bibitem{SkI3} P.-G. Reinhard and H. Flocard, Nucl.\ Phys.\ {\bf A584}, 467 
(1995) .
\bibitem{KDE0v1} B.~K. Agrawal, S. Shlomo, and V. Kim Au,
  Phys.\ Rev.\ C {\bf 72}, 014310 (2005).
\bibitem{LNS} L.~G. Cao, U. Lombardo, C.~W. Shen, and N.~V. Giai,
 Phys.\ Rev.\ C {\bf 73}, 014313 (2006).
\bibitem{SkT123} F. Tondeur, M. Brack, M. Farine, and J.~M. Pearson,
  Nucl.\ Phys.\ {\bf A 420}, 297 (1984).
\bibitem{SV-sym32} P. Kl\'{u}pfel, P.-G. Reinhard, T.~J. B\"urvenich,
  and J. A. Maruhn, Phys.\ Rev.\ C {\bf 79}, 034310 (2009).
  \bibitem{Angeli13} I. Angeli and K.~P. Marinova,
  At.\ Data  Nucl.\ Data Tables {\bf 99}, 69 (2013).
\bibitem{Friar97} J.~L. Friar, J. Martorell, and D.~W.~L. Sprung,
  Phys.\ Rev.\ A {\bf 56}, 4579 (1997).
\bibitem{Goddard13} P.~M. Goddard, P.~D. Stevenson, and A. Rios,
  Phys.\ Rev.\ Lett.\ {\bf 110}, 032503 (2013).
\bibitem{Minomo11} K. Minomo, T. Sumi, M. Kimura, K. Ogata, Y.~R. Shimizu, and M. Yahiro, Phys.\ Rev.\ C {\bf 84},
034602 (2011).
\bibitem{Minomo12} K. Minomo, T. Sumi, M. Kimura, K. Ogata, Y.~R. Shimizu, and M. Yahiro, Phys.\ Rev.\ Lett.\ {\bf 108},
052503 (2012).
\bibitem{Sumi12} T. Sumi, K. Minomo, S. Tagami, M. Kimura, T. Matsumoto, K. Ogata, Y.~R. Shimizu, and M. Yahiro, Phys.\ Rev.\ C {\bf 85}, 
064613 (2012).
\bibitem{Watanabe14} S. Watanabe {\it et al.},
Phys.\ Rev.\ C {\bf 89}, 044610 (2014).
\bibitem{Horiuchi17}
  W. Horiuchi, S. Hatakeyama, S. Ebata, and Y. Suzuki,
Phys. Rev. C {\bf 96}, 024605 (2017).
\bibitem{Oyamatsu03} K. Oyamatsu and K. Iida,
  Prog.\ Theor.\ Phys.\ {\bf 109}, 631 (2003).  
\bibitem{Koura05} H. Koura, T. Tachibana, M. Uno, and M. Yamada, 
Prog.\ Theor.\ Phys.\ {\bf 113}, 305 (2005).
\bibitem{Trippa08}
L. Trippa, G. Col\'o, and E. Vigezzi,
Phys.\ Rev.\ C {\bf 77}, 061304 (2008).
\bibitem{Kolomeitsev17} E.~E. Kolomeitsev, J.~M. Lattimer, A. Ohnishi,
  and I. Tews, arXiv:1611.07133.
\end{thebibliography}
\end{document}